\documentclass[11pt,preprint]{aastex}
\usepackage{longtable,lscape}

\newcommand\mum{\mu{m}}
\def\kmsMpc{\hbox{km} \, \hbox{s}^{-1} \hbox{Mpc}^{-1}}
\def\ewhb{\hbox{EW}(H\beta)}

\begin{document}

\title{Metallicities of Emission-Line Galaxies from HST ACS PEARS and HST WFC3 ERS Grism Spectroscopy at $0.6 < z < 2.4$}
\author{Lifang Xia\altaffilmark{1}, Sangeeta Malhotra\altaffilmark{1}, James Rhoads\altaffilmark{1}, 
Nor Pirzkal\altaffilmark{2}, Amber Straughn\altaffilmark{3}, Steven Finkelstein\altaffilmark{4}, 
Seth Cohen\altaffilmark{1}, Harald Kuntschner\altaffilmark{5}, Martin K\"ummel\altaffilmark{5}, 
Jeremy Walsh\altaffilmark{5}, Rogier A. Windhorst\altaffilmark{1}, and Robert O'Connell\altaffilmark{6}}
\altaffiltext{1}{School of Earth and Space Exploration, Arizona State University, AZ, 85287-1404; lifang.xia@asu.edu}
\altaffiltext{2}{Space Telescope Science Institute, Baltimore, MD 21218, USA}
\altaffiltext{3}{Astrophysics Science Division, Goddard Space Flight Center, Code 665, Greenbelt, MD 20771, USA}
\altaffiltext{4}{George P. and Cynthia Woods Mitchell Institute for
  Fundamental Physics and Astronomy, Department of Physics and Astronomy, Texas A \& M University, College Station, TX 77843, USA}
\altaffiltext{5}{European Southern Observatory, Karl Schwarzschild Str. 2, D 85748 Garching, Germany}
\altaffiltext{6}{Department of Astronomy, University of Virginia, Charlottesville, VA 22904-4325, USA}

\begin{abstract}

Galaxies selected on the basis of their emission line strength show
low metallicities, regardless of their redshifts. We 
conclude this from a sample of faint galaxies at redshifts between $0.6
< z < 2.4$, selected by their prominent emission lines in
low-resolution grism spectra in the optical with the Advanced
Camera for Surveys (ACS) on the Hubble Space Telescope (HST) and in the near-infrared
using Wide-Field Camera 3 (WFC3). Using a sample of 11 emission line galaxies (ELGs) at
$0.6<z<2.4$ with luminosities of $-22\lesssim M_B\lesssim-19$ which have [OII], H$\beta$, and [OIII] line flux
measurements from the combination of two grism spectral surveys, we
use the $R23$ method to derive the gas-phase oxygen abundances:
7.5 $<$12+log(O/H)$<$8.5. The galaxy stellar masses are derived using Bayesian
based Markov Chain Monte Carlo ($\pi$MC$^2$)  fitting of their
Spectral Energy Distribution (SED), and span
the mass range 8.1 $<$ log($M_*/$M$_{\odot}$) $<$ 10.1. These galaxies
show a mass-metallicity (M-L) and Luminosity-Metallicity (L-Z) relation, which is offset by --0.6 dex 
in metallicity at given absolute magnitude and stellar mass relative to
the local SDSS galaxies, as well as continuum selected DEEP2 samples
at similar redshifts. The emission-line selected galaxies most
resemble the local ``green peas'' galaxies and Lyman-alpha
galaxies at $z\simeq0.3$ and $z\simeq2.3$ in the M-Z and L-Z relations and their morphologies.
The $G-M_{20}$ morphology analysis shows that 10 out of 11 show disturbed morphology, even as the star-forming
regions are compact. These galaxies may be intrinsically metal poor, being 
at early stages of formation, or the low metallicities may be due to gas infall and accretion due to mergers.

\end{abstract}

\section{Introduction}

Nebular lines from HII regions are signposts for detection and
measurement of current star-formation. They are also useful for
measuring the metallicity of galaxies. From such studies \citep{lequeux79,garnett87,skillman89,zaritsky94}
we have learned the mass-metallicity and mass-luminosity relations
(e.g. Tremonti et al. 2004), whereby galaxies with higher stellar
mass and higher absolute luminosity show higher metallicities. It is
expected, and observed, that going to higher redshifts should show a shift
in the mass-metallicity relation \citep{erb06,mannucci09}. Higher redshift galaxies do
show a lower metallicity for the same given stellar mass (e.g. Erb et al
2006, Maiolino et al. 2008) for galaxies in the early stages of star-formation. 
Effects of downsizing are also seen in  mass metallicity effects. 
Since lower mass galaxies continue star-formation until later epochs, 
one would expect the slope of the mass-metallicity relation to also change
the offset in the M-Z and L-Z relation. Zahid et al. (2010) show that at $z=0.8$, 
the high mass ($ M > 10^{10.6}M_{\odot}$) galaxies have attained the metallicities seen 
for the same mass galaxies at $z=0$, but low mass galaxies  ($ M \approx 10^{9.2}M_{\odot}$) 
still show a metallicity deficit compared to the same mass galaxies at $z=0$. 

In order to go fainter (and lower stellar masses) at higher redshifts,
we analyze nebular line emission of 11 galaxies in Chandra Deep Field-
South, observed with HST-ACS grism in the optical (from the PEARS program; PI: Malhotra) 
and HST-WFC3 grism (from the ERS program; PI: O'Connell; e.g., Straughn et al. 2011) at near-infrared wavelengths. 
This sample is selected to show emission lines in the slitless spectra, reaching limits of 26.7 mag
and redshifts at $z\lesssim2.3$. Together, these grism data sets span a wavelength range from $\lambda=$0.6--1.6 $\mum$. 
This allows us to measure metallicities using the R23 diagnostic indicator, $R23$ = ([OII]+[OIII])/H$\beta$, \citep{pagel79,kewley02}
for a wide range of redshift, $0.5\lesssim z\lesssim 2.4$, 
without interference by the Earth's atmosphere. Much of this redshift range is inaccessible to 
ground-based observations due to H$_2$O absorption bands, and even more is lost to OH airglow emission lines. 
Our work demonstrates the crucial value of slitless HST spectra in studying the physical properties 
of galaxies at an otherwise challenging redshift range.

The paper is organized as below. In $\S$ 2 we briefly introduce the surveys and the data sample. 
In $\S$ 3 we present the measurements of the metallicity and the stellar mass, and assess the metallicity 
accuracy by comparing with the metallicity measured from follow-up Magellan spectroscopy of two
galaxies. We show the results of the mass-metallicity (M-Z) relation 
and the luminosity-metallicity (L-Z) relation in $\S$ 4. Finally, we discuss the results and give 
our conclusions in $\S$ 5.  We use a ``benchmark'' cosmology with
$\Omega_m = 0.27$, $\Omega_\Lambda = 0.73$, and $H_0 = 71 \kmsMpc$ \citep{komatsu11},
and we adopt AB magnitudes throughout this paper.  

\section{Data}

The HST/ACS G800L Probing Evolution and Reionization Spectroscopically survey 
(PEARS, PI: S. Malhotra, program ID 10530) is the largest survey conducted
to date with the slitless grism spectroscopy mode of the HST Advanced Camera 
for Surveys. PEARS provides low-resolution (R $\sim$ 100) 
slitless grism spectroscopy in the wavelength range from 6000{\AA} to 9700{\AA}. 
The survey covers four ACS pointings in the GOODS-N (Great Observatories Origins
Deep Survey North) field and 
five ACS pointings in the CDF-S (Chandra Deep Field South) fields. Eight of these 
PEARS fields were observed in 20 orbits each (three roll angles per field),
yielding spectra of all objects of AB$_{F850LP}\lesssim26.5$ mag. The ninth field was 
the Hubble Ultra Deep Field (HUDF), which was observed in 40 orbits.  
Combined with the earlier data from the GRAPES program (the GRism ACS Program
for Extragalactic Science; PI: S. Malhotra, program ID 9793), the HUDF field
reaches grism depths of AB$_{F850LP}\lesssim27.5$ mag.

The emission lines most commonly identified from the PEARS grism data are 
[OII]$\lambda$3727{\AA}, the [OIII]$\lambda\lambda$4959,5007{\AA} doublet, and H$\alpha$6563{\AA}. 
Due to the low spectral resolution, the H$\beta$ line is only
marginally resolved from the [OIII] doublet. With the ACS G800L 
grism's wavelength coverage, galaxies at $0.6<z<0.9$ can be
observed in both the [OII] and [OIII] lines, and galaxies at redshifts 
$0.9<z<1.5$ can be observed in only the single line of [OII]$\lambda$3727{\AA},
and at $z<0.5$ in the $H\alpha$ lines of typical line fluxs $\sim1.5$--$2\times10^{-17}$erg$\;$cm$^{-2}\;$s$^{-1}$ \citep{straughn09}.

The HST Wide Field Camera 3 (WFC3) Early Release Science (ERS) (PID GO-11359, PI: O'Connell) program consists of 
one field observed with both the G102 (0.8--1.14 microns; R$\sim$210) and G141 
(1.1--1.6 microns; R$\sim$130) infrared grisms, with two orbits of observation per grism.
This field overlaps with the ACS G800L PEARS grism survey, and hence faint galaxies can be observed with composite spectra 
in the wavelength range from $\lambda\simeq$0.6--1.6 $\mu$m with the detection of the emission lines, such as 
H$\alpha$ at $z\lesssim1.6$, [OIII] doublet at $0.2\lesssim z\lesssim2.4$, and 
[OII] doublet at $0.6\lesssim z\lesssim3.6$ with a S/N $\gtrsim$ 2 line flux 
limit of $\sim3.0\times10^{-17}$erg$\;$cm$^{-2}\;$s$^{-1}$ \citep{straughn11}.

\citet{straughn09} selected 203 emission line galaxies (ELGs) from the
PEARS southern fields, using a 2-dimensional line detection and extraction procedure.  
\citet{straughn11} presented a total catalog of 48 emission-line galaxies from the WFC3 ERS II program \citep{windhorst11}, 
demonstrating the unique capability of the WFC3 to detect star-forming galaxies in the infrared reaching to fluxes of 
AB$_{(F098M)}\lesssim$ 25 mag in a depth of 2 orbits. The combination of these two catalogs yields a sample of 11 ELGs with detection 
of the [OII], [OIII] and H$\beta$ lines in the composite spectral range 0.6--1.6 $\mu$m, which enables us to utilize the R23 method
to measure metallicity, and to extend the study of the mass-metallicity relation of ELGs continuously from $z\simeq$0.6 to 2.4.
We compare the selection of the [OIII] line fluxes, the equivalent width (EW), redshifts, 
and the absolute B-band magnitude of the 11 ELGs from the combined catalog
with respect to the \citet{straughn09} PEARS ELGs sample and the \citet{straughn11} WFC3 ERS ELGs
sample. 

Figure 1 shows the distributions of line fluxes, EWs, redshift and absolute magnitudes of the ERS and PEARS samples,
along with the current sample of 11 galaxies selected as having emission lines in both optical and NIR spectra.
The empty black solid line histogram represents that of the PEARS ELGs sample. The blue 45 degree tilted solid line filled histogram
represents the WFC3 ERS ELGs sample. And the red -45 degree tilted line filled histogram is that of the sample in this paper.
Because the ERS observations are shallower than the PEARS samples, we are confined to relatively bright end of the PEARS sample. 
The top left panel shows that the [OIII] emission line fluxes in our sample are representative 
of the two parent samples at $>5\times10^{-17}$erg$\;$cm$^{-2}\;$s$^{-1}$, which is the bright end of the two parent samples.
The EW([OIII])s (left bottom panel) are in the range from 10s to 1,000s {\AA} with peak at 100s of the parent samples. 
The redshifts (right top panel) and the absolute B-band magnitudes (right bottom panel) are representative of 
the ERS parent sample while offset to high redshift with respect to the PEARS sample, 
which is mainly at $z<1$ and extends to $M_B\sim-17$. 

\begin{figure}[htbp]
\begin{center}
\vspace{-0.5cm}
\figurenum{1} \epsscale{0.9}
\hspace{-1.0cm}
{\plotone{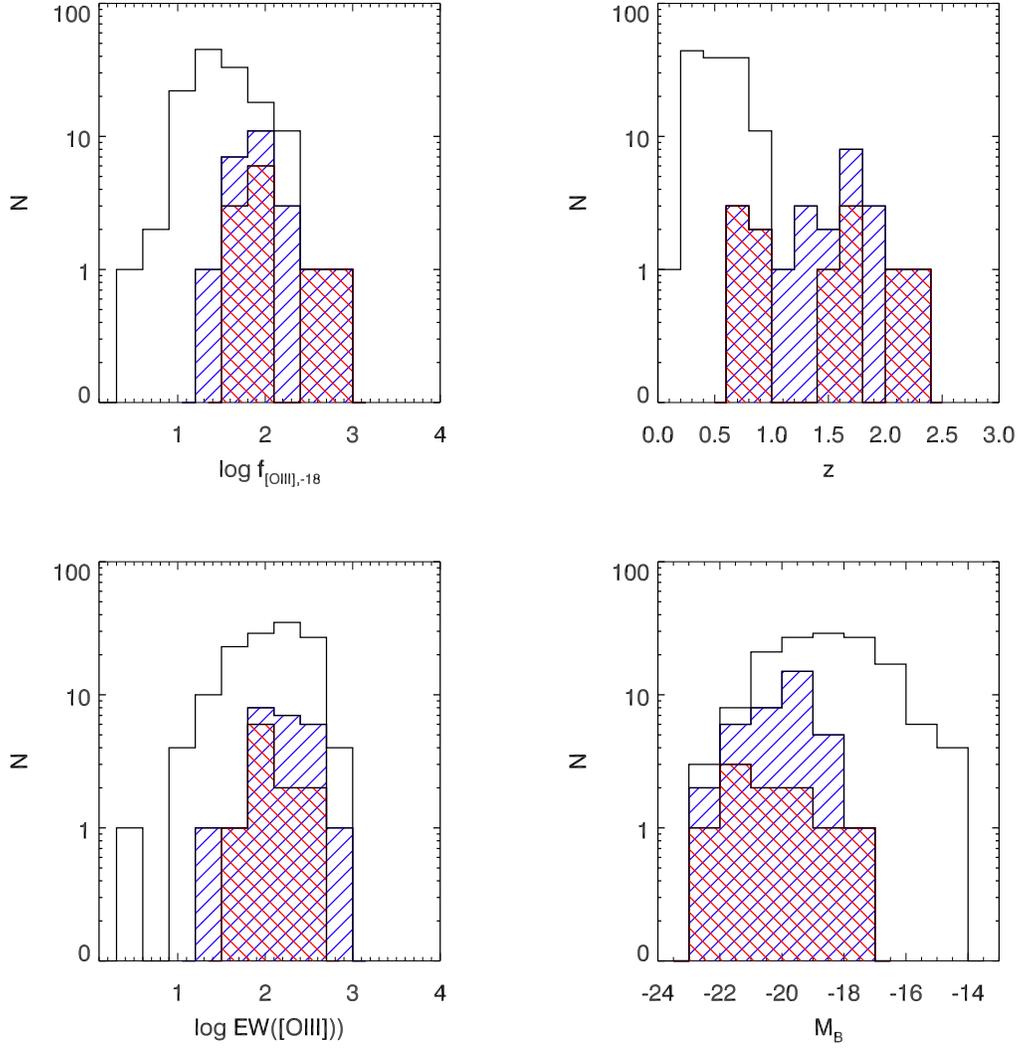}}
\vspace{1.5cm}
\caption{Histogram distributions of the [OIII] line fluxes 
($f_{[OIII],-18}=f_{[OIII]}/10^{-18}$erg$\;$cm$^{-2}\;$s$^{-1}$), the equivalent width (EW), redshifts, 
and the absolute B-band magnitude of the 11 ELGs from the combined catalog (red --45 degree tilted line filled) 
with respect to the \citet{straughn09} PEARS ELGs sample (empty black solid line) and the \citet{straughn11} WFC3 ERS ELGs
sample (blue 45 degree tilted line filled).}
\end{center}
\end{figure}

The HST/ACS PEARS data reduction was similar to the GRAPES project's data
analysis \citep{pirzkal04}, while further steps for identifying
emission line sources are described in \citet{meurer07} and \citet{straughn09}.
The analysis of the WFC3 ERS data is discussed in \citet{windhorst11} and \citet{straughn11}.  
The emission line fluxes are measured from 1D extracted spectra, using
the {\textit IDL} code {\textit mpfit} to fit single or multiple Gaussian line profiles. 
Due to the marginal splitting of the H$\beta$ and [OIII] doublet, the [OIII] line 
is fitted with a double Gaussian profile with the ratio of 
[OIII]$\lambda$4959 to [OIII]$\lambda$5007 constrained to be 1:3 with the same Gaussian widths. 

The H$\beta$ line wavelength is fixed at the redshifted wavelength of 4861 {\AA}, 
given by the observed wavelength of the stronger [OIII]$\lambda$5007 line.
The underlying H$\beta$ absorption amounts are obtained by fitting galaxy SEDs 
(discussed in detail in the next section, Pirzkal et al. 2011)
with the population synthesis model of \citet{bruzual03}.
The EW of the H$\beta$ absorption features range from 4 -- 7 {\AA}, which agrees with the amount obtained
in other studies, e.g. $\sim$ 3$\pm$2 {\AA} \citep{lilly03}. 
The absorption feature is smoothed to the same Gaussian profile as the [OIII] line, and then added to the grism spectra.
The absorption-corrected H$\beta$ line flux is finally measured by adding a Gaussian profile (same as that of 
the [OIII] Gaussian profile) with changing amplitude at the fixed wavelength, on the [OIII] already-fitted double Gaussian profiles.
An H$\beta$ line flux of S/N$>$3 is assumed as detection, and for line fluxes with S/N$<$3 
(1$\sigma\sim5\times10^{-18}$erg$\;$cm$^{-2}\;$s$^{-1}$), we use a 3$\sigma$ upper limit
to the H$\beta$ line flux, which give in a lower limit to the galaxy oxygen abundance on the lower branch (see next section).

The amount of dust extinction is also obtained from the SED fitting, and ranges from $Av=0$--1.2 mag.
The extinction correction is done using the {\textit IDL} code {\textit calz$\_$unred} (written by W. Landsman), based on the reddening curve 
from \citet{calzetti00}. Studies show that the gas can suffer more extinction than the stellar content, hence we assume
E(B-V)$_{stellar}$=0.44E(B-V)$_{gas}$, as has been found locally \citep{calzetti00}. Due to the degeneracy of the extinction
and the stellar population age, the extinction values have large uncertainties. The uncertainties of the extinction values 
are folded into the uncertainties in the 
metallicity. The results show that the uncertainty due to the extinction is in the order of 0.02--0.1 dex, and the dominant part
of the uncertainties in the metallicities result from the faint line flux of H$\beta$ compared to [OIII]$\lambda$5007. 

Table 1 lists the extinction corrected emission line fluxes and restframe equivalent widths of the
[OII]$\lambda$3727, H$\beta$, and the [OIII] doublet for the 11 galaxies in the sample, along with the WFC3 ERS ID and the redshift. 
Figure 2 shows the grism spectra with the Gaussian fit profiles of the [OII]$\lambda$3727, H$\beta$ and [OIII] doublet lines of the 11 galaxies.
Figure 3 shows the GOODS-S $i$-band postage stamps of the 11 galaxies. 

To assess the morphologies of the galaxies in our sample, we measure the Gini coefficient $G$, which quantifies 
the relative distribution of the galaxy's flux,
and the second-order moment of the brightest 20\% of the galaxy's flux \citep{abraham03,lotz04}, M$_{20}$ from the galaxy images. 
Figure 4 shows the distribution of the
galaxies in the G-M$_{20}$ plane with the empirical line dividing normal galaxies with merger/interacting galaxies \citep{lotz04}. 
The blue stars represent that measured from GOODS $B$-band image and the red triangles show that measured from GOODS $i$-band image.
We can see that from the $B$-band image, all of the galaxies lie above the dashed line, which is the region of the outlier galaxies 
showing merger/interacting 
and dwarf/irregular morphologies. From the $i$-band image, 8 out of 11 galaxies are on and above the line and 3 are below the line.
The visual check of the galaxies below the dashed line shows that two galaxies (246, 578) have obvious multiple knots and 
irregular shape, and the galaxy 258 is in the region of dwarf galaxies, which is in agreement with the low mass estimation log$(M)=8.74M_{\odot}$. 
Therefore, we see that 10 out of 11 
show disturbing morphologies, interacting companions and tidal features, which demonstrate the ongoing active star-formation in these galaxies.
At the same time, the half light radii of the galaxies are shown in Table 2, which span the range from 1 -- 8 kpc, with 8 out of 11, 
$r_{1/2}<$ 3 kpc, showing compact morphology.

\begin{figure}[htbp]
\begin{center}
\figurenum{2} \epsscale{0.9}
\hspace{-1.0cm}
{\plotone{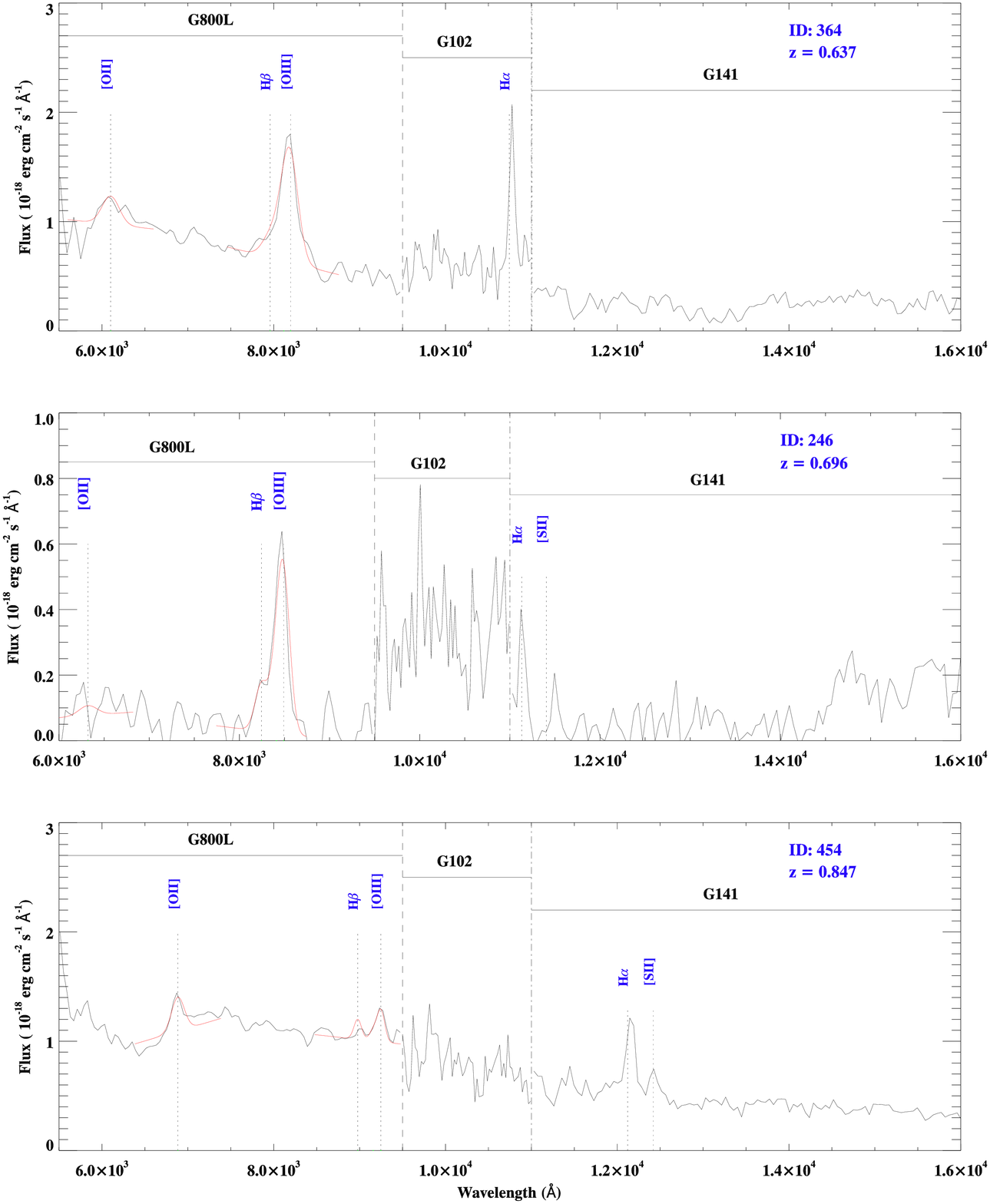}}
\vspace{-1.cm}
\caption{Example of composite grism spectra from the HST/ACS PEARS G800L grism spectroscopy and the HST/WFC3-IR ERS G102 and G141 grism spectroscopy.
The emission lines, [OII]$\lambda$3727, H$\beta$, and [OIII]$\lambda$5007, H$\alpha$ and [SII] are detected.
The H$\beta$, and [OIII] doublet are detected in both G800L and G102 grisms, 
and the G102 grism resolves the [OIII]$\lambda\lambda$4959,5007.
The fitting of the [OIII] doublet is constrained to make the ratio of the [OIII]$\lambda$4959 to [OIII]$\lambda$5007
fluxes 1:3, and to use the same line width for both. 
The detection of both [OII] and [OIII] in the composite spectra enables 
the meatallicity measurement using the R23 method.}
\end{center}
\end{figure}

\begin{figure}[htbp]
\begin{center}
\figurenum{3} \epsscale{0.9}
{\plotone{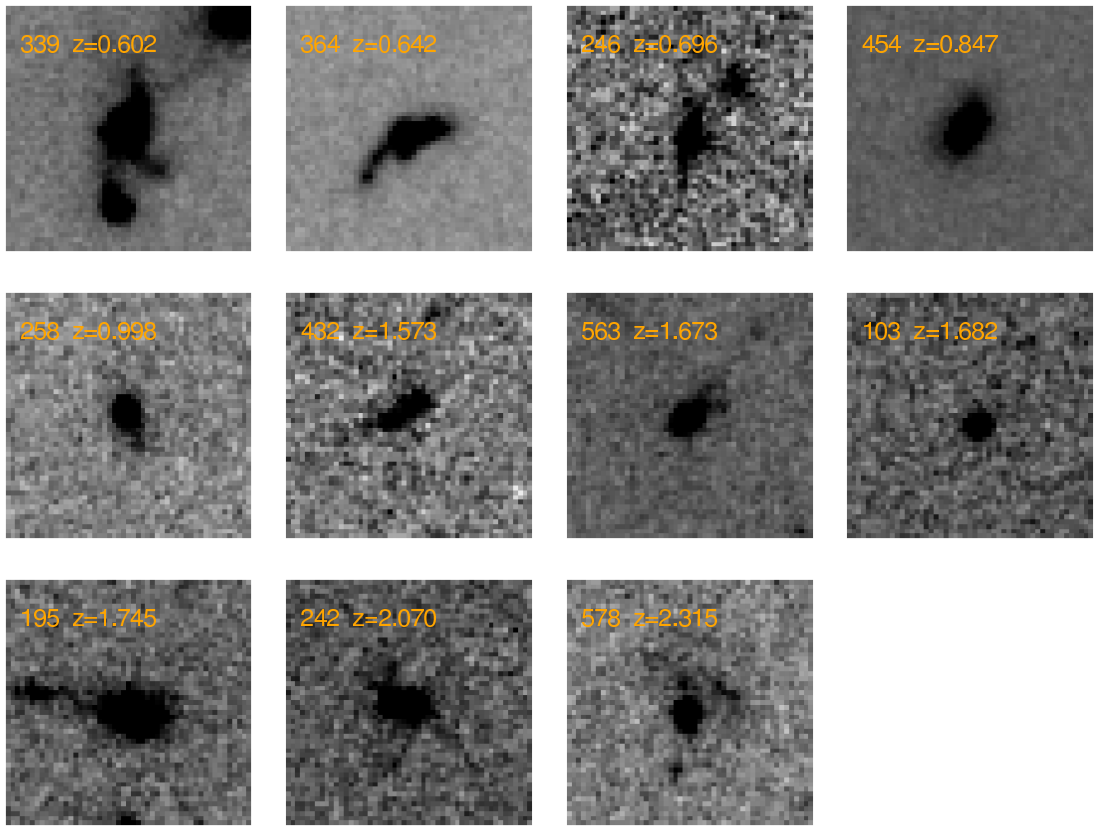}}
\caption{The GOODS-S $i$-band postage stamps of the 11 ERS galaxies in our sample. The irregular morphologies, interacting companions, and
tidal features demonstrate ongoing star-formation of these galaxies.}
\end{center}
\end{figure}

\begin{figure}[htbp]
\begin{center}
\figurenum{4} \epsscale{0.9}
{\plotone{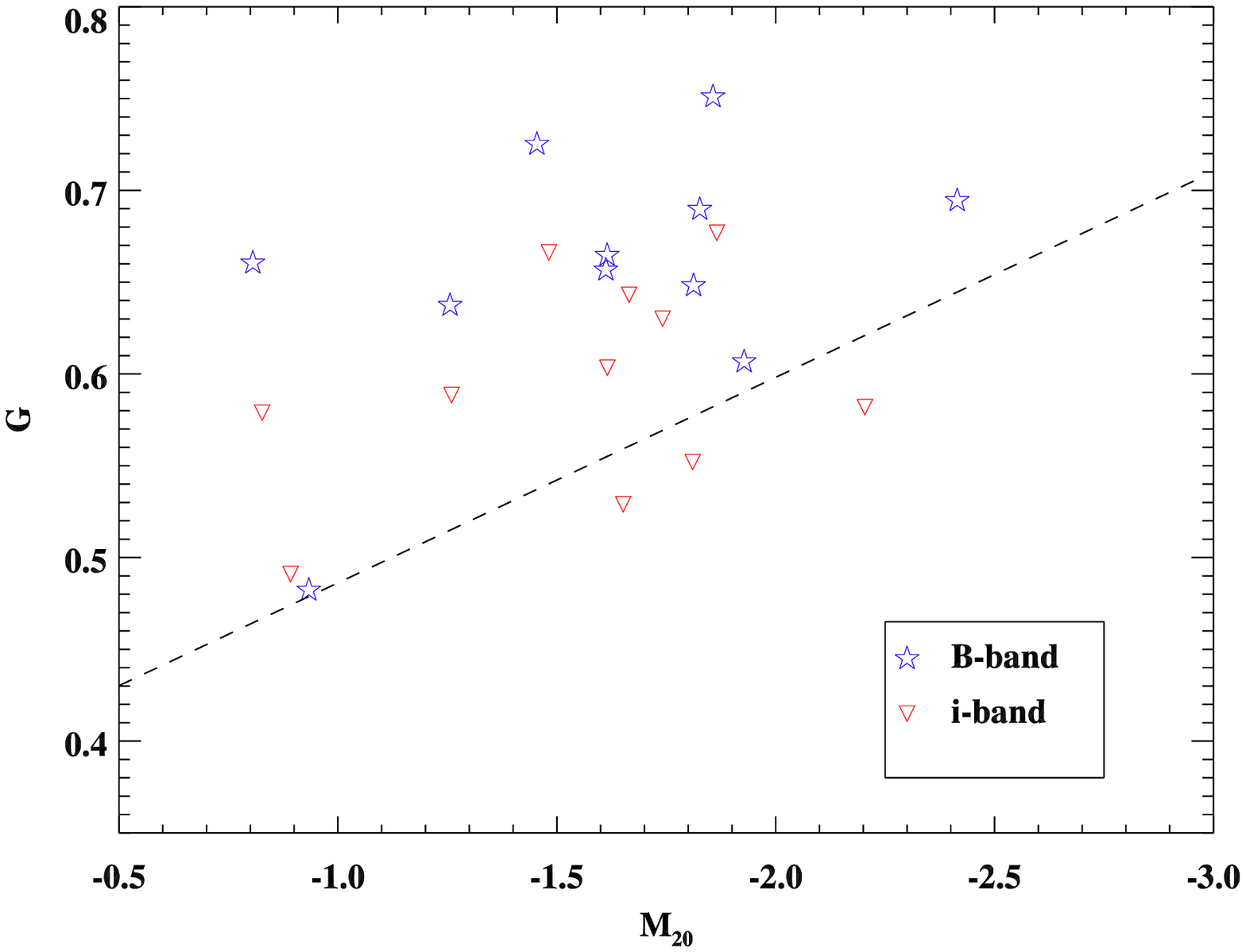}}
\caption[The morphology analysis, G vs. M$_{20}$.]{Gini coefficient $G$ vs. M$_{20}$ to demonstrate the morphology analysis of the 11 galaxies in our sample. 
The dashed line is the empirical line dividing interacting galaxies (upper region) with normal galaxies (lower region) from \citet{lotz04}. The blue stars
represent galaxies based on $B$-band image analysis. The red triangles are that based on $i$-band image analysis. Most galaxies lie above the line demonstrating
disturbed morphologies.}
\end{center}
\end{figure}

\section{Measurements}

\subsection{Metallicity}

Using the strong nebular lines [OII]$\lambda$3727, H$\beta$, and the [OIII] doublet measured from the combined grism spectra,
we measure the gas-phase oxygen abundance by the most commonly used $R23$ ($R23$ = ([OII]+[OIII])/H$\beta$) diagnostic indicator \citep{pagel79,kewley02}.  
We calculate the metallicities by iteration, using the 
parameterized calibrations between the oxygen abundance 12+log(O/H), the ionization parameter $q$, and $R23$ that are derived from theoretical 
photoionization models by Kewley \& Dopita (2002) and Kobulnicky \& Kewley (2004).

We select the R23 method, because it relies on measuring some of the brightest
nebular emission lines, which allows it to be used for faint galaxies in the
distant universe.  However, it has one major drawback, which
is that the relation between $R_{23} \equiv (f_{OII} + f_{OIII}) / f_{H\beta}$
and the gas phase oxygen abundance $12 + log(O/H)$ is in general double-valued,
with both a high- ($12 + log(O/H)>8.5$) and a low- ($12 + log(O/H)<8.5$), metallicity branch solution. For the present
data set, we rely on a set of three secondary metallicity indicators to
decide whether our galaxies lie on the upper or lower branch.  
First is the ``O32'' ratio, $f_{OIII} / f_{OII}$.  While this is primarily 
sensitive to the ionization parameter $q$ \citep{kewley02}, it can also be used
as a branch indicator \citep{maiolino08}, with ratios of $f_{OIII} / f_{OII} > 2$ indicating 
a lower branch solution, and $f_{OIII} / f_{OII} < 1$ indicating an
upper branch solution. Second is the ratio $f_{OIII} / f_{H\beta}$, with
$f_{OIII} / f_{H\beta} > 3$ indicating $7.4 \lesssim 12 + log(O/H) \lesssim 8.5$ 
\citep{maiolino08}.  Third is the equivalent width of the $H\beta$ line.  
\citet{hu09} show that $\ewhb$ correlates with metallicity, such that
$\ewhb \gtrsim 30$\AA\ implies a lower branch solution, and $\ewhb \lesssim 10$\AA\ 
implies the upper branch solution. 

Other popular branch indicators --- notably the [OIII]$\lambda$4363
line strength and the $N2$ diagnostic indicator ($N2$ = log
([NII]$\lambda$6584/H$\alpha$) --- are not practical for our data set,
given the faintness of the [OIII]$\lambda$4363 line, and the blending
of [NII]$\lambda$6584 with H$\alpha$ in HST grism spectroscopy.
Nevertheless, the combination of $\hbox{EW}(H\beta)$,
$f_{OIII}/f_{OII}$, and $f_{OIII}/f_{H\beta}$ provides reasonable
confidence in our branch identifications for most of our sample.

Figure 5 shows the $\log(R23)$ versus 12+log(O/H) for our 11 ELGs on the lower
branch. The lines represent the model relationships between $\log(R23)$ and 12+log(O/H)
at two ionizations with $q=1.0\times10^7,1.0\times10^8$. The use of the upper limit of H$\beta$
line fluxes gives the lower limit of R23, and thereafter the lower limit of the metallicities
on the lower branch, which are shown as right arrows and upward arrows.
Since the galaxies are put on the lower branch, 
Table 2 shows $\log(R23)$, the ionization parameter $\log(q)$, and the oxygen abundances and
their corresponding uncertainties. 
The large uncertainties on the oxygen abundances are mainly due to the large 
fractional flux uncertainties for H$\beta$ in our data. 
All of the galaxies are on the lower branch, and some are near the
peak in the $\log(R23)$ vs. metallicity curve, where the branch indicators 
become both ambiguous and largely irrelevant, and their metallicities are near 12+log(O/H)=8.5.

The galaxy oxygen abundances in our sample span the range from $7.5<12$+log(O/H$)<8.5$, i.e, $\sim$ 0.1 Z$\odot$ -- Z$\odot$. 
(A solar metallicity has Z$\odot$=0.015 and 12+log(O/H) = 8.72, see Allende Prieto et al. 2001). 
As we see from table 2, the low redshift galaxies at $0.6<z<1$ have an average metallicity of 12+log(O/H)$\simeq$7.95, 
and the galaxies at $z>1$ have higher average metallicity of 12+log(O/H)$\simeq$8.26,
brighter absolute mangitudes and larger stellar masses (see Table 2). This shows the selection effects at low redshift and high
redshift of the sample. At same magnitude and line flux limits, the galaxies selected with larger redshifts tend to be more massive,
brighter and higher metallicity galaxies. Hence, to evaluate the evolution of the metallicity for same mass galaxies at different
redshifts, we need to enlarge the sample to include faint low-mass galaxies at high redshift.

Two galaxies out of the 11 ELGs (ERS ID numbers 339, 364) have followup Magellan spectroscopy, which covers the wavelength range 
from 4000 to 9000 {\AA}, with a wavelength-resolution of $\sim$ 3 {\AA} \citep{xia12}.
The metallicities measured from the Magellan spectra using the R23 method on the strong emission lines [OII], H$\beta$ and [OIII] doublet
give 12+log(O/H) = 8.07 $\pm$ 0.14 for ERS339 and 8.18 $\pm$ 0.15 for ERS364 \citep{xia12}.
The metallicities obtained from the HST ACS/WFC3 grism spectra (12+log(O/H) = 8.10$^{+0.20}_{-0.16}$ for ERS339 and 8.22$^{+0.16}_{-0.13}$ for ERS364) 
and that obtained from the Magellan spectroscopic spectra agree to within 1 $\sigma$ ($\sim$ 0.1 dex), underscoring the feasibility of 
emission-line galaxy metallicity measurements using the HST/WFC3 IR grism data.

\begin{figure}[htbp]
\begin{center}
\figurenum{5} \epsscale{0.9}
{\plotone{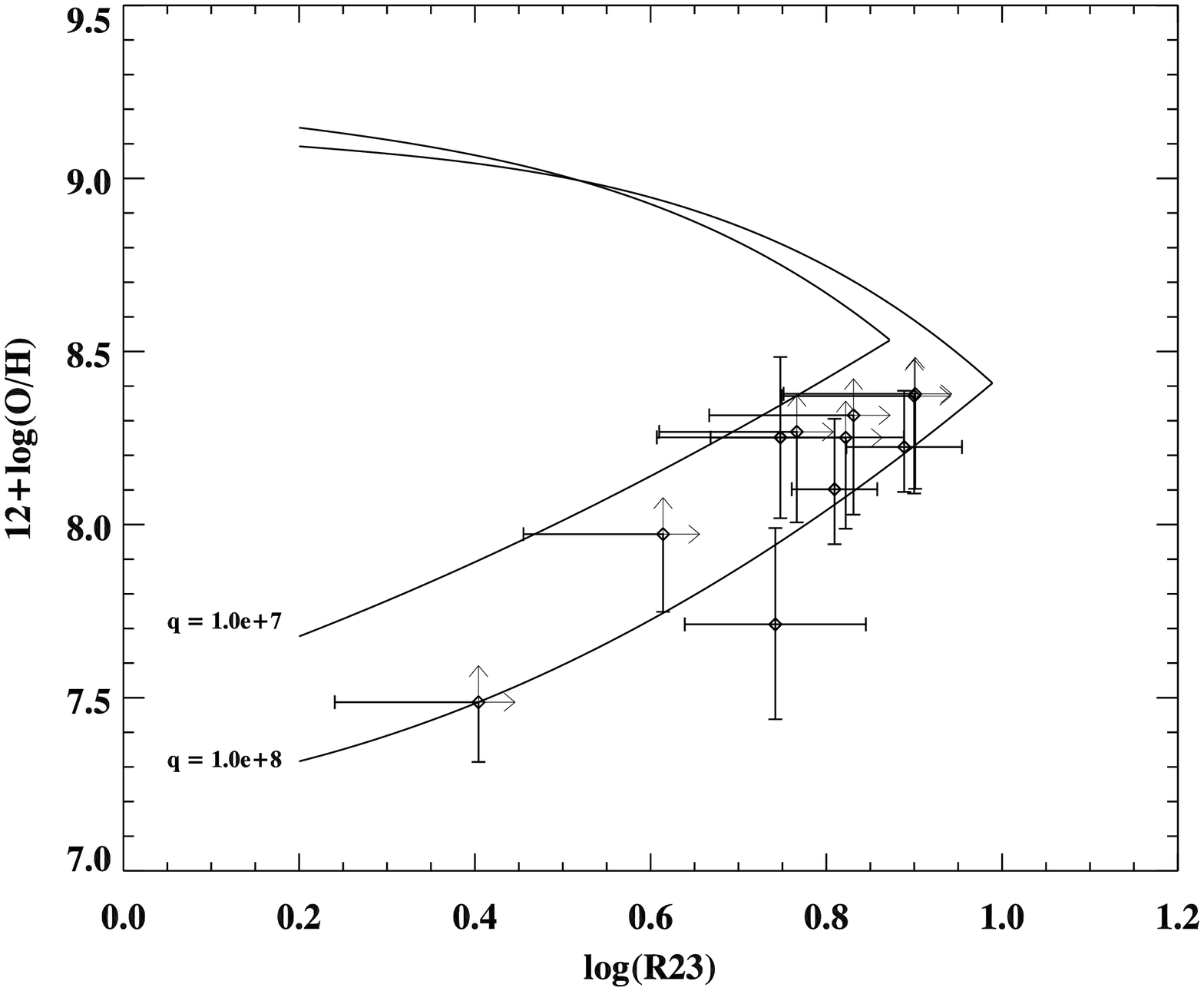}}
\caption{The log(R23) versus oxygen abundance for the 11 ELGs in our sample. The overplotted lines represent the theoretical lines at $q=1.0\times10^7,
1.0\times10^8$
\citep{kobulnicky04}. All of the galaxies are put on the lower branch according to the branch criteria. The 3$\sigma$ upper limit of the H$\beta$ 
line fluxes give the lower limit of log(R23) and hence the lower limit of 12+log(O/H) at the lower branch, which are shown as arrows.}
\end{center}
\end{figure}

\subsection{Stellar Mass}

The galaxy stellar masses are derived by comparing the observed photometry with the model spectra library produced by the 
\citet{bruzual03} stellar population synthesis code (BC03, hearafter). 
The galaxies in our sample are located in the ACS pointings of the GOODS-South field. The optical broadband 
$BViz$ photometry is obtained from HST/ACS GOODS version 2.0 images \citep{giavalisco04}. 
The UV photometry in F225W, F275W, and F336W, as well as the near-IR photometry in F098M (Ys ), F125W (J), and F160W (H)
are from the new WFC3 ERS mosaics \citep{windhorst11}. 
In this paper, we adopt the galaxy stellar masses measured by the method of Bayesian based Markov Chain Monte Carlo ($\pi$MC$^2$),
which allows us to compare the observations to arbitrarily complex models, and to compute 95\% credible intervals that provide 
robust constraints for the model parameter (see Pirzkal et al. 2011 for details). 
The models are generated using the single (SSP), two 
(SSP2) stellar instantaneous populations, or an exponentially decaying star formation history model (EXP).
The parameters assumed in the models are Salpeter initial mass function (IMF), metallicities ranging from $Z=0.004$ to 0.02 (Z$_{\odot}$), 
the stellar population ages, the relative ratio between the old and young stellar populations,
the \citet{calzetti00} extinction law, and the half-life $\tau$ value in the case of EXP 
models. 
We divide by a factor of 1.8 to make the galaxy stellar masses derived from Salpeter IMF in this paper consistent with 
that in other studies derived from Chabrier (2003) IMF \citep{erb06}.
The comparison of the R23 and the SED best-fit metallicities shows agreement to 
within $\sim$ 0.1 dex for 6 galaxies, and $\sim$ 0.5 dex higher metallicity for the other 5 galaxies from the SED fitting method. 
The higher metallicity has very little effect on the estimated galaxy masses as shown by \citet{pirzkal11}. 
The results of the galaxy stellar masses and stellar population ages 
are shown in the sixth column of Table 2. The galaxies show young ages of 20--90 Myr and low masses $\sim$
$10^{8}-10^{10}M_{\odot}$.

\section{Results}

The wide spectral coverage of the HST/ACS PEARS and WFC3 ERS composite grism spectra provide galaxies at $0.6<z<2.4$ with full set of emission
lines [OII], H$\beta$ and [OIII], which extend the study of the evolution of the L-Z relation and the M-Z relation to redshift
$z\simeq$ 2.5. In this section, we will show the results of the luminosity-metallicity relation and the mass-metallicity relation,
which provide important clues to the evolution of galaxies by comparing with the relations at different redshifts.

\subsection{L-Z relation}

Previous results show important evolution of the slope and the zero point of the L-Z relation with respect to redshift, 
decreasing metallicity with increasing redshift at a given luminosity.
With the sample of 11 grism ELGs at $0.6<z<2.4$, we investigate the evolution of the L-Z relation with redshift. 
Following traditions, we present the rest-frame absolute $B$-band magnitude as a measure of the luminosities. 
The restframe $B$-band absolute magnitudes are computed from the best-fit SED with the BC03 stellar population synthesis model.

Figure 6 shows the relationship between the absolute rest-frame $B$ magnitude versus the gas-phase oxygen abundance derived from $R23$ 
diagnostic indicator. The lines plotted in Figure 6 are the local L-Z relation 
obtained by \citet{tremonti04} for $\sim$53,000 SDSS galaxies at $z\sim0.1$ (solid line), the L-Z relation obtained by \citet{zahid11}
from 1350 DEEP2 emission line galaxies at z $\sim$ 0.8 (dashed line), that obtained by \citet{hu09} from a sample of Ultra-Strong
Emission-Line (USELs) galaxies at $z\simeq$0--1 (dotted line and empty stars), and that of \citet{salzer09} for 
15 star-forming galaxies at $z\sim0.3$ (open upside down triangles). 
Our sample of 11 galaxies span a range in luminosity --17 $<M_B<$ --23 and in metallicity $7.5<12$+log(O/H$)<8.5$. 
The red solid dots represent the galaxies with $z>1$, and the green triangles represent the galaxies with $z<1$.
The blue solid line shows the best linear fit of the 11 galaxies, a relation of 12+log(O/H$)=(4.75\pm0.86)-(0.17\pm0.04)M_B$
with a correlation coefficience of --0.77.

Compared to the other relationships shown in Figure 6, ACS+WFC3 grism galaxies are about 7 magnitudes brighter in luminosities
than the local SDSS galaxies and the $z\sim$ 0.8 DEEP2 galaxies at fixed metallicity. 
The DEEP2 sample \citep{zahid11} shows little evolution compared to the SDSS sample, about $\sim$ 0.1 dex 
relative to the local L-Z relation, while the ERS grism galaxies show $\sim$ 0.6 dex lower metallicities 
than the SDSS galaxies at given luminosity. 
The grism galaxies show a good match with metal-poor galaxies of \citet{hu09,salzer09} along the fitted L-Z relation.

The \citet{hu09} USELS galaxies have high equivalent width with EW(H$\beta>30${\AA}), extend to 
fainter galaxies to M$_B\sim$ --16 and show low metallicities of $7.1<12$+log(O/H$)<8.4$. The 
\citet{salzer09} are [OIII]-selected galaxies ([OIII]/H$\beta>$3) at $z\sim$0.3 and show brighter luminosity
and higher metallicities. The difference of the galaxies on the L-Z figure shows the different physical 
properties of the three samples: the USELS are basically selected to be fainter dwarf galaxies, the low redshift
\citet{salzer09} are [OIII]-selected brighter galaxies. Since the three samples follow
well of the L-Z relationship of the metal-poor galaxies, and the L-Z relations of the SDSS galaxies and 
the DEEP2 galaxies are obtained by averaging large samples, we conclude that the big offset in the L-Z relation 
between the local and the three metal-poor galaxies samples is due to the selection of 
a sample of young strong emission-line star-forming galaxies, which will be further illustrated in the next subsection.


\begin{figure}[htbp]
\begin{center}
\figurenum{6} \epsscale{1.}
\hspace{-1.0cm}
{\plotone{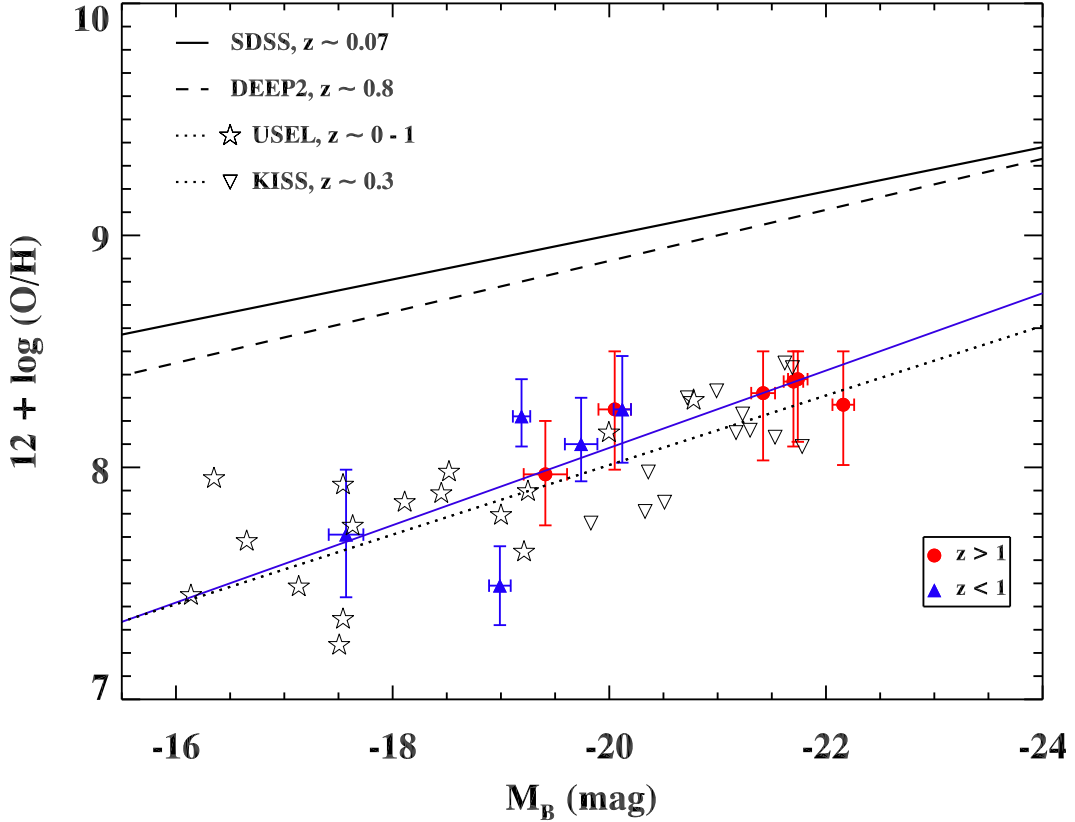}}
\caption{$L$-$Z$ relation between the rest-frame $B$-band absolute magnitude versus the oxygen abundance for the 10 emission line galaxies at $0.6<z<2.4$.
The metallicity is derived from the $R23$ indicator and the $x$-axis is the rest-frame $B$-band absolute magnitude. 
The red solid dots represent the galaxies with $z>1$, and the blue triangles represent the galaxies with $z<1$. 
The solid line represents the relation obtained by \citet{tremonti04} for SDSS star-forming galaxies at $z\sim0.1$. 
The dashed line illustrates the relation obtained by \citet{zahid11} for DEEP2 galaxies at $z\sim0.8$. 
The dotted line and the empty stars show the relation obtained by \citet{hu09} for USEL galaxies at $z=0-1$. 
The empty upside down triangles are that of \citet{salzer09} for [OIII]-selected $z\sim0.3$ galaxies.
The blue solid line shows the best linear fit of our sample, which gives a relationship 
of $12+$log$(O/H)=(4.75\pm0.86)-(0.17\pm0.04)M_B$.
The PEARS sample shows an offset by about --0.6 dex in metallicity relative to the local relation at $z\sim0.1$.}

\end{center}
\end{figure}

\subsection{M-Z relation}

Figure 7 shows the relation between the stellar masses and the gas-phase oxygen abundances for the 11 star-forming galaxies in our sample at $0.6<z<2.4$.
The solid line represents the $M$-$Z$ relation at $z\sim0.1$ from \citet{tremonti04} for the local SDSS galaxies, which are selected to be
star-forming galaxies based on lines H$\alpha$, H$\beta$, and [NII].
The dashed line shows the $M$-$Z$ relation at $z\sim0.8$ for the 1350 H$\beta$ selected blue DEEP2 galaxies from \citet{zahid11}. 
The dotted line and the dash-dotted line are UV-color selected galaxies 
at $z\sim2.3$ from \citet{erb06} and the UV-selected Lyman Break Galaxies (LBGs) at $z\sim3.1$ from \citet{mannucci09}, respectively. 
The red line shows the best fit to the M-Z relation for the ELGs in our sample.

The green triangles illustrate the sample of the ``green peas'' from Cardamone et al. (2009) and 
Amorin et al. (2010), which are extremely compact (r$<3$ kpc) star-forming galaxies at $0.11<z<0.35$ selected by color from the SDSS 
spectroscopic observation, with an unsual large equivalent width of up to $\sim$ 1000 {\AA}. 
We recalculate the gas-phase oxygen metallicity by the R23 method for the ``green peas'' sample.
Also plotted are the Ly$\alpha$ emitters at $z\sim0.3$, and $\sim$2.3 from Finkelstein et al. (2011a,b), shown in empty red circles and black asterisks 
with 2 $\sigma$ and 3 $\sigma$ upper limits, and one extremely metal poor galaxies XMPG WISP5-230 \citep{atek11}. 
All data presented have been scaled to a Chabrier (2003) IMF. To ensure the consistency of the comparison, the conversion given by
\citet{kewley08} is used to convert to the same metallicity calibration of \citet{kobulnicky04} to avoid the differences arising from different 
metallicity indicators \citep{zahid11}. The metallicity of the XMPG galaxy from \citet{atek11} is measured by the direct $Te$ method, and is not converted 
to the same metallicity diagnostic due to the absence of the [OII] flux
and the conversion relationship between the direct $Te$ method and the R23 method in \citet{kobulnicky04}.

From Figure 7, our grism galaxies span the range $8.1<$log$(M_{*}/$M$_{\odot})<10.1$ and $7.5<12$+log(O/H$)<8.5$, with the average 
values of $<$log(M$_{*}/M_{\odot})>=9.3$ and $<$12+log(O/H)$>=8.1$. Although this is a small sample, it shows a similar correlation 
between metallicity and stellar mass, increasing oxygen abundance with the increase of the stellar masses. 
The red dots in Figure 7 show the 6 galaxies with redshift $z>1$ and with emission lines observed in WFC3 ERS. The blue triangles represent the galaxies
with $0.6<z<1$. We fit the mass-metallicity relation with a second-order polynomial \citep{maiolino08}: 
\begin{equation}
12+log(O/H)=A[log(M)-log(M_0)]^2+K_0,
\end{equation}
the best fit parameters to the 11 ELGs in our sample give A=-0.07, log(M$_0$)=11.87, K$_0$=8.63. 
From Table 2, we see that these high redshift galaxies have higher stellar masses with a mean of 
$<logM_{*}/$M$_{\odot}>\simeq$ 9.6 and higher metallicities with a mean of $<$12+log(O/H)$>$$\simeq$ 8.3.
The low redshift subsample have lower galaxy stellar masses with a mean of $<$log$M_{*}/$M$_{\odot}$$>$$\simeq$ 8.8 and lower metallicities with a 
mean of $<$12+log(O/H)$>$$\simeq$ 8.0. The offset shown between the high redshift subsample and the low redshfit subsample includes the evolution
of the M-Z relation with redshift, and the selection effect, that for the same emission line detection the high redshift galaxies tend to be more luminous, 
more massive and more metal-enriched than the low redshift galaxies. 

We examine the M-Z relation by comparing our sample with that at different redshift ranges.
Compared with the local relation at $z\sim0.1$, the SDSS galaxies with comparable stellar mass to the average of 
the grism sample, $M_{*}\sim10^{9.3}$ M$_{\odot}$, have $12+$log$(O/H)\simeq$ 8.8, which is about $\sim$ 0.6 dex higher than the average of the grism galaxies.
For the low redshift subsample with a mean of $z\simeq0.8$, the M-Z relation show a large offset of $\sim$ 0.6 dex with that of \citet{zahid11}
at $z\simeq0.8$ too. This big difference between our sample and that of \citet{tremonti04} and \citet{zahid11} is mainly due to the different 
selection criteria of the galaxies. The local SDSS galaxies \citep{tremonti04} and the DEEP2 galaxies \citep{zahid11} 
are obtained from large spectroscopic survey, and the M-Z relations show the average relationships of the dominant galaxy populations
at that redshift. Table 3 lists the physical properties including redshift range, selection, absolute magnitude, emission line EW,
half light radius and SFR of the different comparing samples. We can see that the SDSS and DEEP2 samples are not selecting high EW 
star-forming galaxies compared with the ``green peas'' \citep{amorin10}, USELS \citep{hu09}, LBGs \citep{mannucci09} and our PEARS/ERS
ELGs, which are biased to high EW emission-lines (up to $\sim$ 1000 {\AA}) and compact ($r_{1/2}<3$ kpc) galaxies.

For the high-redshift subsample with a mean of $z\simeq2$, the M-Z relation shows an offset of $\sim$ 0.2 dex with respect to that of the LBGs at $z\simeq2.3$ \citep{erb06}. 
The low metallicity galaxies basically fall between the relation at $z\simeq2.3$ and $z\simeq3.1$ and have low metallicities down to 12+log(O/H)$\sim$7.5, 7.7.
The ``green peas'' \citep{hoopes07,overzier08,amorin10} at $z\simeq0.3$ are found to be metal-poor by 
$\sim$ 0.5 dex relative to other galaxies of similar stellar mass, and show compact and distrubed morphology.
From Figure 7, we find that 7 out of 11 of the HST/ACS+WFC3 grism emission line galaxies are in the similar metallicity range 12+log(O/H)$\sim$8.3 and four galaxies
are more metal-poor by up to 0.6 dex, compared with the green peas at the same galaxy stellar masses, which shows significant chemical
enrichment from $z\simeq0.8$ to $z\simeq0.3$ at the low stellar mass range. To confirm this evolution with higher statistical 
significance, we will need larger sample of galaxies extending to low mass faint galaxies at high redshifts.
The strong emission line selected Ly$\alpha$ galaxies at $z\simeq0.3$, at $z\simeq2.3$ and XMPG WISP5$\_$230 at $z\simeq0.7$ show similar lower
metallicities at $7.2<12+log(O/H)<8.2$ with respect to the average M-Z relations obtained from large survey samples.

The detailed analysis of the morphologies, sizes, colors, SSFRs based on the M-Z relation \citep{pirzkal06,xia12} show that the 
strong emission-line selected grism galaxies are biased towards young compact interacting dwarf 
star-forming galaxies. \citet{pirzkal06} shows small physical sizes of $\sim$ 1--2 kpc for the emission line galaxies 
observed from the GRAPES survey, and \citet{xia12} presents high SSFRs $10^{-9}-10^{-7}/yr$ for the ELGs from the PEARS survey.
Since the galaxies in our sample are partly the subsample of the PEARS ELGs, the results of the sizes and the SSFRs are consistent with the
previous results, with $r_{1/2}<3 kpc$ and $SSFR\sim10^{-9}/yr$. This confirms the selection effects of the young compact disturbed
emission line galaxies in the sample.
The early stage of galaxy evolution (downsizing effect) or interaction-induced pristine gas inflow 
picture may account for the offset of the grism galaxies in metallicity relative to the local sample.

\begin{figure}[htbp]
\begin{center}
\figurenum{7} \epsscale{0.9}
\hspace{-1.0cm}
{\plotone{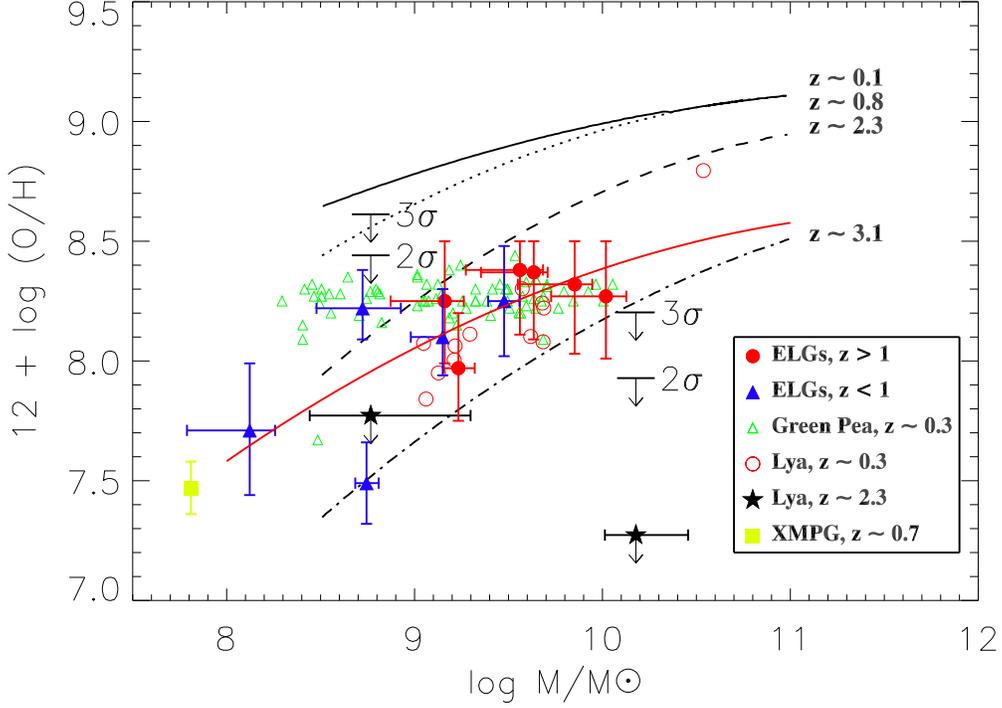}}
\caption{Relation between the stellar masses and the gas-phase oxygen abundances for our sample of 11 ELGs from PEARS and ERS 
grism data at $0.6<z<2.4$. The metallicities are estimated from the $R23$ method. 
The stellar masses are estimated from the SED fitting with $BC03$ model
and have been scaled to a Chabrier (2003) IMF to be consistent with other studies. 
The definition of the points of our sample are the same as Figure 5. Also plotted for comparison are the green peas (empty green triangulars) at z$\simeq$0.3
\citep{amorin10}, Ly$\alpha$ galaxies at $z\simeq0.3$ and $z\simeq2.3$ \citep{finkelstein11a,finkelstein11b}, and the WISP XMPG galaxy at $z\simeq0.7$ \citep{atek11}.
The solid red line is the best fit of the M-Z relation to the 11 ELGs in our sample. 
The solid line represents the M-Z relation at $z\simeq0.1$ from \citet{tremonti04} for the local SDSS galaxies. 
The dashed line shows the M-Z relation at $z\simeq0.8$ for the 1350 DEEP2 galaxies from \citet{zahid11}. 
The dotted line and the dash-dotted line are that at $z\simeq2.3$ from \citet{erb06} and at $z\simeq3.1$ from \citet{mannucci09}, respectively. 
The M-Z relations at different redshifts are calibrated to the same metallicity indicator of \citet{kobulnicky04} from \citet{zahid11}. 
The large offset of $\sim$ 0.5 dex of this sample relative to the other relations at similar redshift demonstrates that these galaxies may be
at the early-stages of galaxy evolution. Infall of gas due to mergers is another popular explanation, e.g. \citet{peeples09}.} 
\end{center}
\end{figure}

\section{Discussion and Summary}

We use a sample of 11 emission line galaxies at $0.6<z<2.4$ observed by HST/ACS PEARS and HST WFC3 ERS programs
at 0.6--1.6 $\mu$m to demonstrate the effectiveness of the grism spectra (R $\sim$ 100--300) used for the metallicity measurement.
With the [OII], H$\beta$, and [OIII] lines in the composite spectra of the two grism spectra surveys, we use the $R23$ method 
to derive the gas-phase oxygen abundances, 12+log(O/H). For two galaxies which have the follow-up Magellan 
spectroscopy, the metallicities obtained from the grism spectra and from the Magellan spectroscopic spectra are consistent to within
1 $\sigma$ (0.1 dex), which demonstrates the feasibility of the HST/WFC3 IR grism spectra used here for the study of galaxy metallicities. 

The measured gas-phase abundances are in the low metallicity range $7.5<12$+log(O/H$)<8.5$. 
The galaxy stellar masses are derived from MCMC SED fitting and span the range 8.1 $<$ log($M_*/$M$_{\odot}$) $<$ 10.1. 
Both the L-Z relation and the M-Z relation show that with the increase of the galaxy stellar mass or the luminosity, 
the metallicity increases, which agrees with the enrichment history of galaxy evolution. 
The M-Z relation of this sample show significant offset by about --0.6 dex in metallicity at given stellar mass
relative to the local M-Z relation from SDSS galaxies and the galaxies from the DEEP2 survey at similar redshifts $z\simeq0.8$. 
The L-Z relation is fitted by a straight line of 12+log(O/H$)=(4.75\pm0.86)-(0.17\pm0.04)M_B$ 
with a correlation coefficience of --0.77,
which is also offset by about --0.6 dex in metallicity relative to the local and $z\simeq0.8$ L-Z relations. 

Our sample of galaxies at $z\simeq0.8$ show similarity to the local green peas in morphology and low metallicity. 
Two galaxies show significant poorer metallicity by $\sim$0.5 dex compared with the ``green peas'' at the same galaxy stellar masses, 
which signifies different physical processes in the galaxy evolution and chemical enrichment from $z\simeq0.8$ to $z\simeq0.2$ 
at the low stellar mass range. 
The different contribution by downsizing and gas inflow/outflow need to be examined in detail by larger samples further.

By comparing the PEARS/ERS sample with other emission-line selected samples 
\citep{erb06,hu09,mannucci09,salzer09,amorin10,zahid11,finkelstein11a,finkelstein11b} in Table 3, 
we find that the physical properties of the ELGs galaxies at different redshifts show great similarities: e.g. 
(1) ultra-strong emission lines of about $10^{-17}$erg$\;$cm$^{-2}\;$s$^{-1}$, high emission-line EWs up to $\sim$ 1000 {\AA}, 
and hence very high SSFRs to $10^{-9}/yr$;
(2) compact morphology ($r_{1/2}<3kpc$); (3) evidence for mergers/interactions from the asymmetries of the morphology, 
such as companions and wispy tidal tails around a compact star-forming region, 
three ``green peas'' shown in Cardomone et al. (2009), and 10 out of 11 galaxies in our sample (see Figure 3). 
Hence, we conclude that the offsets shown in the M-Z and L-Z relations with respect to that obtained from average of large sample 
are mainly due to the selection effects based on prominent emission lines. 
\citet{van11} shows an abundant population of extreme emission line galaxies (EELGs) from the HST/WFC3 CANDELS Survey 
(Cosmic Assembly Near-IR Deep Extragalactic Legacy Survey) and confirms the physical properties of low stellar masses $\sim$
10$^8M_\odot$, and strong outflows due to enormous starbursts in the EELGs by the HST/WFC3 grism spectra. 

Taken together, the properties of the ELGs: compact starbursts, low metallicities, disturbed morphologies, and low masses,
indicate that these are dwarf galaxies undergoing their early stages of galaxy evolution with prominent signs of strong activities
of interaction (gas accretion and outflow) with companion galaxies. 
Both the downsizing effect and inflow/outflow play important roles in these low metallicity galaxies' evolution. 
To examine the mode of the star-formation of these low-mass, low-metallicity galaxies in the whole scenario of galaxy
evolution requires a larger sample of this kind of ELGs from optical to NIR spectroscopy with 
morphologies to give us a more comprehensive picture of these galaxies. \citet{trump11} presents a sample of 28 emission
line galaxies at $z\sim2$ with prominent [OIII] and H$\beta$ in the GOODS-S region of the Cosmic Assembly Near-infrared Deep 
Extra-galactic Legacy Survey(CANDELS). Combined with the PEARS spectra, this sample will greatly enhance the sample at redshift
$z~2$ at the low-mass low-metallicity region of the M-Z relation, which is important to study and understand 
the physical processes effecting galaxy evolution.

\begin{acknowledgements}

We thank the anonymous referee for the comments which are very helpful in clarifying and improving this paper. 
This paper is based on Early Release Science observations made by the WFC3 Scientific Oversight Committee. 
PEARS is an HST Treasury Program 10530 (PI: Malhotra). Support for program was provided
by NASA through a grant from the Space Telescope Science Institute, which is operated by
the Association of Universities for Research in Astronomy, Inc., under NASA contract
NASA5-26555 and is supported by HST grant 10530.

\end{acknowledgements}



\begin{landscape}
\begin{center}

\begin{table}[ht]
\scriptsize
\caption[measurements]{The extinction corrected emission line fluxes and equivalent widths of the PEARS/ERS grism galaxies. The H$\beta$ line fluxes are absorption corrected by the SED fitting. The detections of the H$\beta$ line are set with S/N$>$3. The 3$\sigma$ upper limit of the H$\beta$ line is used for galaxies with S/N$<$3. These galaxies are marked with stars. } \label{} 
\begin{tabular}{ccccccccccc}
\\
\hline \hline \\[-2ex]

 ID & $z$ & R.A. & DEC. &  E(B-V)   & [OII]3727  & EW([OII]) & H$\beta$ & EW(H$\beta$)   & [OIII]  & EW([OIII]) 
\\[0.5ex]
 &  & (deg) & (deg) & (mag) & ($10^{-18}$erg$\;$s$^{-1}\;$cm$^{-2}$) & ({\AA}) & ($10^{-18}$erg$\;$s$^{-1}\;$cm$^{-2}$) & (\AA)
& ($10^{-18}$erg$\;$s$^{-1}\;$cm$^{-2}$) & ({\AA}) 
\\[0.5ex] \hline

  339 &   0.602 &  53.0773392 & -27.7081985 &   0.30$^{+0.30}_{-0.30}$    &   645.51 $\pm$  162.65  &    29 &  468.41 $\pm$   45.19   &   61 & 2373.95 $\pm$   56.24  &   334  \\       
  364 &   0.642 &  53.0693359 & -27.7090893 &   0.03$^{+0.18}_{-0.03}$    &    80.90 $\pm$   15.93  &    40 &   50.35 $\pm$    7.24   &   38 &  308.61 $\pm$    9.59  &   248  \\       
  246 &   0.696 &  53.0700035 & -27.7165890 &   0.03$^{+0.16}_{-0.03}$    &     4.50 $\pm$    4.50  &    26 &   22.90 $\pm$    5.22   &  352 &  121.91 $\pm$    6.92  &  1605  \\       
  454 &   0.847 &  53.0761719 & -27.7011452 &   0.16$^{+0.09}_{-0.10}$    &   166.57 $\pm$   17.80  &    28 &   45.22 $\pm$   13.92   &   11 &   86.35 $\pm$   18.02  &    22  \\            
  258 &   0.998 &  53.0857124 & -27.7113400 &   0.03$^{+0.06}_{-0.03}$    &    29.98 $\pm$    4.25  &    74 &   73.63 $\pm$   35.74 $^\star$   &  525 &  241.91 $\pm$   47.48  &   729  \\       
  432 &   1.573 &  53.0484200 & -27.7095337 &   0.08$^{+0.16}_{-0.08}$    &   101.97 $\pm$   23.19  &    44 &   24.21 $\pm$   11.76 $^\star$   &   16 &  132.11 $\pm$   15.56  &   108  \\       
  563 &   1.673 &  53.0705452 & -27.6956444 &   0.14$^{+0.20}_{-0.14}$    &    93.91 $\pm$   17.34  &    46 &   13.95 $\pm$    9.06 $^\star$   &   19 &  122.04 $\pm$   11.86  &   165  \\       
  103 &   1.682 &  53.0633392 & -27.7272835 &   0.06$^{+0.13}_{-0.06}$    &    43.55 $\pm$   10.23  &    93 &    9.84 $\pm$    7.81 $^\star$   &   45 &   52.83 $\pm$   10.33  &   193  \\           
  195 &   1.745 &  53.0656700 & -27.7203941 &   0.09$^{+0.09}_{-0.09}$    &    87.84 $\pm$   13.89  &    37 &   21.25 $\pm$    8.28 $^\star$   &   17 &  109.87 $\pm$   10.91  &    94  \\           
  242 &   2.070 &  53.0821304 & -27.7137547 &   0.19$^{+0.17}_{-0.17}$    &    94.79 $\pm$   29.03  &    72 &   13.39 $\pm$    8.57 $^\star$   &   25 &   79.46 $\pm$   11.19  &   143  \\           
  578 &   2.315 &  53.0589218 & -27.6978111 &   0.26$^{+0.11}_{-0.19}$    &   116.58 $\pm$   21.06  &    98 &   12.29 $\pm$   10.42 $^\star$   &   10 &   65.98 $\pm$   13.53  &    35  \\

\hline

\end{tabular}

\normalsize

\end{table}
\end{center}
\end{landscape}




\begin{landscape}
\begin{table}[ht]
\scriptsize


\caption[Redshifts and fluxes]{The ionization parameter, metallicity, half-light radius, absolute magnitude, galaxy stellar mass and SFR, SSFR 
of the PEARS/ERS grism galaxies. The missing upper errors in log$R23$ and 12+log(O/H) denote the lower limits due to the use of the upper limits 
of H$\beta$ line fluxes. The galaxy stellar masses derived from Salpeter IMF are scaled to that from Chabrier (2003) IMF to be consistent in 
comparison with other studies.} \label{} 
\begin{center}
\begin{tabular}{ccccccccccc}

\\[0.5ex]

\hline\hline \\[-2ex]
\multicolumn{1}{c}{ID} & \multicolumn{1}{c}{$z$} & \multicolumn{1}{c}{log($R23)$} & \multicolumn{1}{c}{log $q$}  & \multicolumn{1}{c}{12+log(O/H)} & $r_{1/2}$
& \multicolumn{1}{c}{$M_B$} &  \multicolumn{1}{c}{log $M_*$}  &   Age  &  SFR   &    SSFR   \\[0.1ex]
(1) & (2) & (3) & (4) & (5) & (kpc) & (mag) &  ($M_\odot$) &   (Myr)   &   ($M_\odot$/yr) &  ($\times10^{-9}$/yr)
\\[0.5ex] \hline
\\[-1.8ex]

       339  &  0.604 &   0.81$^{+0.05}_{-0.05}$  &   8.12  $\pm$   1.28  &   8.10$^{+0.20}_{-0.16}$  &   8.10    &    --19.74  &   9.19 $^{+0.02}_{-0.34}$    &  --- &  15.20   $\pm$   1.72    &  10.64  $\pm$   4.16    \\  
       364  &  0.637 &   0.89$^{+0.07}_{-0.07}$  &   7.97  $\pm$   0.10  &   8.22$^{+0.16}_{-0.13}$  &   1.72    &    --19.19  &   8.72 $^{+0.21}_{-0.25}$    & 56.9$_{ -53.7}^{ +27.7}$  & 1.70   $\pm$   0.39    &  1.90  $\pm$   0.84    \\  
       246  &  0.691 &   0.74$^{+0.10}_{-0.10}$  &   8.58  $\pm$   0.51  &   7.71$^{+0.28}_{-0.27}$  &   1.97    &    --17.57  &   8.12 $^{+0.14}_{-0.33}$    & 50.4$_{ -47.1}^{ +0.04}$  & 1.06   $\pm$   0.21    &  3.31  $\pm$   0.81    \\  
       454  &  0.847 &   0.75$^{+0.14}_{-0.14}$  &   7.29  $\pm$   0.17  &   8.25$^{+0.23}_{-0.23}$  &   1.37    &    --20.12  &   9.48 $^{+0.08}_{-0.08}$    & 60.8$_{ -32.1}^{ +71.5}$  & 2.41   $\pm$   1.40    &  0.60  $\pm$   0.37    \\  
       258  &  0.997 &   0.40$^{+}_{-0.16}$     &   8.02  $\pm$   0.11  &   7.49$^{+}_{-0.17}$     &     2.20    &  --18.99  &   8.74 $^{+0.06}_{-0.06}$    &  90.1$_{ -35.3}^ {+35.8}$ &  5.78  $\pm$    3.66  &    7.83 $\pm$    5.61  \\    
       432  &  1.573 &   0.82$^{+}_{-0.15}$     &   7.58  $\pm$   0.13  &   8.25$^{+}_{-0.26}$     &     4.14    &  --20.05  &   9.16 $^{+0.10}_{-0.29}$    &  51.0$_{ -47.8}^{ +44.5}$ &  2.71  $\pm$    2.44  &    1.73 $\pm$    1.56  \\    
       563  &  1.673 &   0.90$^{+}_{-0.15}$     &   7.64  $\pm$   0.13  &   8.37$^{+}_{-0.28}$     &     1.58    &  --21.70  &   9.63 $^{+0.07}_{-0.28}$    &  46.5$_{ -43.3}^ {+52.2}$ &  3.66  $\pm$    2.24  &    0.51 $\pm$    0.37  \\    
       103  &  1.682 &   0.61$^{+}_{-0.16}$     &   7.46  $\pm$   0.12  &   7.97$^{+}_{-0.22}$     &     1.07    &  --19.41  &   9.23 $^{+0.08}_{-0.07}$    &  93.1$_{ -68.8}^{ +72.8}$ &  1.59  $\pm$    1.35  &    0.88 $\pm$    0.83   \\   
       195  &  1.745 &   0.90$^{+}_{-0.15}$     &   7.61  $\pm$   0.10  &   8.38$^{+}_{-0.27}$     &     1.89    &  --21.74  &   9.56 $^{+0.13}_{-0.29}$    &  23.9$_{ -20.8}^{ +29.2}$ &  3.80  $\pm$    2.02  &    2.30 $\pm$    1.22   \\   
       242  &  2.070 &   0.83$^{+}_{-0.16}$     &   7.46  $\pm$   0.21  &   8.32$^{+}_{-0.29}$     &     2.09    &  --21.42  &   9.85 $^{+0.09}_{-0.30}$    &  39.7$_{ -36.5}^{ +55.1}$ & 14.51  $\pm$    9.60  &    1.91 $\pm$    1.37   \\   
       578  &  2.315 &   0.77$^{+}_{-0.16}$     &   7.32  $\pm$   0.16  &   8.27$^{+}_{-0.26}$     &     5.82    &  --22.16  &  10.02 $^{+0.11}_{-0.29}$    &  19.0$_{ -15.8}^{ +40.2}$ & 19.17  $\pm$   12.81  &    0.50 $\pm$    0.53    \\[0.3ex]
\hline

\normalsize

\end{tabular}
\end{center}

\end{table}
\end{landscape}




\begin{landscape}
\begin{center}

\begin{table}[ht]
\scriptsize
\caption[comparison]{The selection criteria and physical properties of the comparison samples in the paper.} \label{} 
\begin{tabular}{lcccccccc}
\\
\hline \hline \\[-2ex]

Sample & Survey & $z$ & Selection & $f_{line}$ & $M_B$ & EW   & r$_{1/2}$  & SFR \\[0.1ex]
(1) & (2) & (3) & (4) & ($10^{-17}$erg$\;$cm$^{-2}\;$s$^{-1}$) & (mag) & ({\AA}) & (kpc) & (M$_{\odot}$/yr) 
\\[0.5ex] \hline

Tremonti et al. (2004);         & SDSS         & 0.005$<z<$0.25 & H$\alpha$, H$\beta$, [NII]  & ---        & (-16, -22)     & EW(H$\alpha)\sim$3-200       & --         & --       \\
Salzer et al.(2009);            & KISS         & $z\sim$0.3     & [OIII]                      & ---        & (-19.5, -22.5) & ---                          & --         & --       \\
Finkelstein et al. (2011b); LAEs& HETDEX       & $z\sim$0.3     & Ly$\alpha$                  & $>$ 1400   & ---            & EW([Ly$\alpha$]) $\sim$ 20-40& --         & --       \\
Amorin et al. (2010); green pea & SDSS         & 0.11$<z<$0.35  & color                       & ---        & ---            & EW([OIII]) $<$ 1000          & $<$ 3      & $<$ 30   \\
Zahid et al. (2011)             & DEEP2        & 0.75$<z<$0.82  & H$\beta$, color             & ---        & (-19.5, -22)   & $<$EW(H$\beta)>\sim$8.9      & --         & --       \\
Hu et al. (2009); USELS         & DEMOS        & 0$<z<$1        & [OIII], H$\alpha$           & $>$ 1.5    & (-16, -21)     & EW(H$\beta)<$ 500            & --         & --       \\
Erb et al. (2006);              & LRIS-B       & $z\sim$2.3     & UV-colors                   & $>$ 15     & (-20.5, -23.5) & ---                          & --         & 20 -- 60 \\
Finkelstein et al. (2011a); LAEs& HETDEX       & $z\sim$2.3     & Ly$\alpha$                  & $>$ 200    & ---            & EW([OIII]) $<$ 300           & --         & 20 -- 40 \\
Mannucci et al. (2009); LBGs    & AMAZE, LSD   & 2.6$<z<$3.4    & ---                         & $>$ 1.1    & ---            & ---                          & 0.7 -- 2.4 & 5 -- 40  \\
This paper                      & PEARS, ERS   & 0.6$<z<$2.3    & Emission lines              & $>$ 5      & (-17.5, -22.5) & EW([OIII]) $<$ 1600          & 1 -- 8.1   & 1 -- 20  \\

\hline

\end{tabular}

\normalsize

\end{table}
\end{center}
\end{landscape}


\end{document}